\documentstyle[preprint,aps,tighten]{revtex}
\newcommand{\be}{\begin{equation}}
\newcommand{\ee}{\end{equation}}
\def\ba{\begin{array}}
\def\ea{\end{array}}
\def\bea{\begin{eqnarray}}
\def\eea{\end{eqnarray}}
\def\l{\left}
\def\r{\right}
\begin{document}
\preprint{
\begin{tabular}{r}
IMSc/2000/03/12\\
UWThPh-2000-15\\
March 2000
\end{tabular}
}
\draft 
\title{Neutrino survival probabilities in magnetic fields}
\author
{K.R.S. Balaji\footnote{balaji@imsc.ernet.in}}         
\address
{Institute of Mathematical Sciences,\\ 
Chennai 600$\:$113, India}
\author
{Walter Grimus\footnote{grimus@doppler.thp.univie.ac.at}}
\address
{Institute for Theoretical Physics, University of Vienna,\\
Boltzmanngasse 5, A--1090 Vienna, Austria} 
\maketitle
\begin{abstract}
We show that, for Majorana neutrinos propagating in a constant
magnetic field, the flavour survival probabilities for left-handed
neutrinos is the same as for right-handed neutrinos, i.e., 
$P^M (\nu_{\alpha L}\to\nu_{\alpha L}) =
 P^M (\nu_{\alpha R}\to\nu_{\alpha R})$, where $\alpha = e, \mu, \tau$, 
whereas in the Dirac case the
corresponding probabilities $P^D (\nu_{\alpha L}\to\nu_{\alpha L})$
and $P^D (\bar\nu_{\alpha R}\to\bar\nu_{\alpha R})$
are in general different. This might lead to a novel way to search for
the nature of neutrinos.
We also discuss how this relation for Majorana neutrinos gets modified
when the magnetic field is not constant. However, if matter effects
become important the relation does not hold anymore.
\end{abstract}

\section{INTRODUCTION}

In the past few years neutrino physics has generated tremendous
interest and has been looked upon as a window to physics beyond the standard
model. Recent results from Super-Kamiokande \cite{sk} have boosted this
search for new physics. Both the solar
and atmospheric neutrino anomalies agree well with the neutrino
oscillation hypothesis \cite{oschyp} (for fits to the solar neutrino
data see Ref.\cite{fit}), which requires neutrinos to
be massive. For recent reviews see, e.g., Ref.\cite{review}.
However, it should be kept in mind that neutrino flavour
oscillations is not the only possibility to describe the data. 
It may be that massive neutrinos have non-zero
magnetic moments (MM) and electric dipole moments (EDM), which
could play an important role in the solar neutrino puzzle if
large magnetic fields are present in the interior of the sun \cite{MM}.
In this framework, the attractive scenario of resonant spin -- flavour
transitions \cite{flip} has received attention and good fits to the solar
neutrino data have been obtained (for recent fits see Refs. 
\cite{MMfit} and for reviews see Ref.\cite{MMreview}).
In solutions of the solar neutrino puzzle with non-zero MMs and EDMs,
it is assumed that the solar neutrinos interact with
the magnetic field in the sun to produce right-handed 
neutrinos due to a helicity flip. Right-handed neutrinos are sterile
in the case of Dirac neutrinos and behave like antineutrinos in the
Majorana case. In the latter case, a flavour transition
simultaneous with the helicity flip produces the suppression in solar
neutrino detection with elastic neutrino -- electron scattering.

So far, it is not known if neutrinos possess MMs and/or EDMs. 
The most stringent laboratory bounds come from elastic
neutrino -- electron scattering: for $\bar\nu_e$ reactor neutrinos the
limit is $1.8 \times 10^{-10} \mu_B$ \cite{derbin} whereas for 
$\stackrel{\scriptscriptstyle (-)}{\nu}_{\hskip-3pt \mu}$
the limit is $7.4 \times 10^{-10} \mu_B$ \cite{krakauer}, where
$\mu_B$ is the Bohr magneton. More
stringent limits are obtained with astrophysical considerations
\cite{raffelt}. For a collection of references on limits on neutrino
MMs see also Ref.\cite{caso}. 

In this paper we focus on neutrino survival probabilities.
Apart from the three active
neutrino flavours ($\nu_\alpha$ with $\alpha = e, \mu, \tau$),
we allow an arbitrary number of additional neutrinos
of the sterile type ($\nu_s$) \cite{sterile}. 
In vacuum, the survival probabilities of
left-handed and right-handed (anti)neutrinos are equal:
\be\label{vacuum}
P^D (\nu_{\alpha L}\to\nu_{\alpha L}) =
P^D (\bar\nu_{\alpha R}\to\bar\nu_{\alpha R})
\quad \mbox{and} \quad
P^M (\nu_{\alpha L}\to\nu_{\alpha L}) =
P^M (\nu_{\alpha R}\to\nu_{\alpha R}) \,,
\ee
where the superscripts $D$ and $M$ refer to Dirac and Majorana
neutrinos, respectively. These equalities are not valid if matter effects
\cite{matter} become important, 
but matter effects in neutrino oscillations do not
distinguish between the Dirac and Majorana nature \cite{langacker}.
Here we concentrate
on the situation that matter effects are negligible but the MM and EDM
interaction of neutrinos with magnetic fields becomes important. Thus our
discussion does not apply to solar neutrinos. We
will show that for constant magnetic fields the equality (\ref{vacuum})
remains valid for Majorana neutrinos whereas in general it gets lost 
for Dirac neutrinos. This is to be contrasted with the matter effects.
We will also consider a situation where the
magnetic field depends on $\vec x$ where this equality for Majorana
neutrinos persists.
Though our discussion will be rather formal we hope that it can shed
some light on a possible distinction between Dirac and Majorana
neutrinos. Up to now such efforts are mainly concentrated on
neutrinoless double beta decay \cite{beta}. 

Our paper is organized as follows.
In Section II, we discuss the general formalism for an oscillating
neutrino (both, left and right-handed neutrinos) propagating in an
external electromagnetic field, in the absence of matter effects, and
prove the equality (\ref{vacuum}) for Majorana neutrinos in a constant
magnetic field. In Section III we study some conditions where
equality (\ref{vacuum}) also holds for Dirac neutrinos, but we show
that it is violated in the general Dirac case.
In Section IV we consider non-constant magnetic fields and in Section V
we summarize our results.

\section{GENERAL FORMALISM}

In the following we will work with $n$ neutrino flavours or types. 
We will take into account neutrino mixing and electromagnetic
interactions through MMs and EDMs but we will not consider matter effects.

\subsection{DIRAC CASE}

Let us assume that the neutrino MMs, EDMs and transition MMs and EDMs
are given for the neutrino mass eigenfields $\nu_j$. Then the MM and
EDM interaction of the Dirac neutrinos is expressed by the
Hamiltonian density
\begin{equation}\label{Hem}
{\mathcal{H}}^D_{\mathrm{em}} = 
\frac{1}{2} \bar\nu (\mu + i d \gamma_5) \sigma_{\alpha\beta}
F^{\alpha\beta} \nu 
\quad \mbox{with} \quad \mu^\dagger = \mu, \; d^\dagger = d
\end{equation}
being the magnetic moment and electric dipole moment
matrices, respectively. $F^{\alpha\beta}$ is the antisymmetric electromagnetic 
field tensor. Whereas $\nu_j$ denotes the fields in the mass basis, we
denote the chiral fields in the flavour basis where the charged lepton
matrix is diagonal by $\nu_L$, $\nu_R$. 
The mass matrix $M$ in the neutrino mass term
\be
- {\mathcal{L}}^D_m = \bar\nu_R M \nu_L + \mbox{h.c.}
\ee
is bidiagonalized by 
\be
U_R^\dagger M U_L = \hat{M}
\ee
such that
\be
\nu_{\alpha L} = \sum_j U_{L\, \alpha j}\, \nu_{jL}
\quad \mbox{and} \quad
\nu_{\alpha R} = \sum_j U_{R\, \alpha j}\, \nu_{jR} \,,
\ee
where the indices $\alpha$ denote the neutrino flavours or types. The
diagonalizing matrix $U_L$ is the usual neutrino mixing matrix.

For massive Dirac neutrinos with mixing, the Hamiltonian in the 
\emph{flavour basis} describing the neutrino interacting with a magnetic field 
is given as \cite{flip,minakata,grimus} 
\be\label{HDnu}
H^D_\nu = \l(\ba{cc} \frac{1}{2E_\nu}U_L\hat M^2 U_L^\dagger & 
 -B_+U_L(\mu+id)U_R^\dagger\\
-B_-U_R(\mu-id)U_L^\dagger & \frac{1}{2E_\nu}U_R\hat M^2 U_R^\dagger \ea
 \r) \,,
\ee
where the upper half of the matrix $H^D_\nu$
corresponds to negative helicity while the lower half
corresponds to positive helicity. Assuming that the
neutrino is propagating along the $z$ direction, the magnetic fields $B_\pm$
are defined as 
\be\label{B}
B_\pm = B_x \pm i B_y = B e^{\pm i \beta} 
\quad \mbox{with} \quad B = \sqrt{B_x^2 + B_y^2} \,. 
\ee
In the approximation we are working with, the longitudinal magnetic
field is negligible. The neutrino energy is denoted by $E_\nu$.
Note that for the right-handed Dirac neutrino states there is
no preferred flavour basis because these states do not couple to the
charged leptons. 
The Hamiltonian matrix $H^D_{\bar\nu}$ for Dirac antineutrinos is
obtained by making the replacements
\be\label{nubar}
U_{L,R} \to U^*_{L,R} \,, \quad \mu \to -\mu^T = -\mu^* \,,
\quad d \to -d^T = -d^*
\ee
in the matrix (\ref{HDnu}). The superscript $*$ on the MM and EDM
matrices indicates complex conjugation of all elements of the matrices.
The upper half of $H^D_{\bar\nu}$
maps onto positive and the lower half onto negative helicity.

The Hamiltonian (\ref{HDnu}) is re-expressed in the
mass basis as $\tilde H_\nu$, where
\be\label{Hnutilde}
\tilde H_\nu = \l(\ba{cc} \frac{1}{2E_\nu}\hat M^2 & 
 -B_+(\mu+id)\\
-B_-(\mu-id) & \frac{1}{2E_\nu}\hat M^2 \ea
 \r) \,,
\ee
and the corresponding Hamiltonian for antineutrinos is given by 
\be
\tilde H_{\bar\nu} = \left( \begin{array}{cc} 
\frac{1}{2E_\nu}\hat M^2 & B_+(\mu^*+id^*) \\
B_-(\mu^*-id^*) & \frac{1}{2E_\nu}\hat M^2 
\end{array} \right) \,.
\ee

In order to obtain the survival probabilities of neutrinos
(which have negative helicity) and antineutrinos (which have positive
helicity), one needs to diagonalize the matrices $\tilde H_\nu$ and 
$\tilde H_{\bar\nu}$, respectively. Observing that
\be\label{J}
J^\dagger \tilde H_{\bar\nu} J = \tilde H_\nu^*
\quad \mbox{with} \quad J = \left(
\begin{array}{rr} 0 & \mathbf{1} \\ -\mathbf{1} & 0 
\end{array} \right) \,,
\ee
where $\mathbf{1}$ is the $n \times n$ unit matrix,
we find that $H^D_\nu$ and $H^D_{\bar\nu}$ have the same eigenvalues 
$E_1, \ldots, E_{2n}$. Furthermore, there are unitary matrices
$W_\nu$ and $W_{\bar\nu}$ such that
\be\label{W}
W_\nu^\dagger \tilde H_\nu W_\nu = 
W_{\bar\nu}^\dagger \tilde H_{\bar\nu} W_{\bar\nu} =
{\mathrm{diag}} (E_1,E_2,...,E_{2n}) \,,
\ee
which, according to Eq.(\ref{J}), are related by
\be
W_{\bar\nu} = J W_\nu^* \,.
\ee

The survival probabilities for
Dirac neutrinos (helicity = $-1$) and Dirac antineutrinos (helicity = $+1$)
are then given as
\be\label{PDnu}
P^D (\nu_{\alpha L} \rightarrow \nu_{\alpha L}) =
\left| \, \sum_{j =1}^{2n} 
\left| U_{\alpha j}^\nu \right|^2 e^{-iE_jL} \, \right|^2 
\ee
and 
\be\label{PDnubar}
P^D (\bar\nu_{\alpha R} \rightarrow \bar\nu_{\alpha R}) =
\left| \, \sum_{j =1}^{2n} 
\left| U_{\alpha j}^{\bar\nu} \right|^2 e^{-iE_jL} \, \right|^2 
\,,
\ee
respectively.
In the above probability expressions, $L$ is the distance between
neutrino source and detector. From the $2n \times 2n$ 
matrices $U^ \nu$ and $U^ {\bar\nu}$ diagonalizing $H^D_\nu$ and
$H^D_{\bar\nu}$, respectively, we only need the first $n$ lines 
labelled by the neutrino flavours or types, which
can be expressed by $U_L$ and $W \equiv W_\nu$: 
\be\label{U}
\begin{array}{rll}
1 \leq j \leq n : &
U^\nu_{\alpha j} = \sum_{k=1}^n U_{L\, \alpha k} W_{kj} \,, &
U^{\bar\nu}_{\alpha j} = 
\sum_{k=1}^n U^*_{L\, \alpha k} W^*_{n+k\,j} \,,\\[1mm]
n+1 \leq j \leq 2n : &
U^\nu_{\alpha j} = \sum_{k=1}^n U_{L\, \alpha k} W_{k\, n+j} \,,&
U^{\bar\nu}_{\alpha j} = 
\sum_{k=1}^n U^*_{L\, \alpha k} W^*_{n+k\,n+j} \,.
\end{array}
\ee

\subsection{MAJORANA CASE}

For Majorana neutrinos we start with the same electromagnetic Hamiltonian
density (\ref{Hem}) as in the Dirac case, except that the factor $1/2$ is
replaced by $1/4$ to account for the charge conjugation property
$(\nu_j)^c = \nu_j$ of the Majorana fields.
Then the Hamiltonian $H_M$, corresponding to
$H^D_\nu$, is given by 
\be\label{HM}
H_M = \l(\ba{cc} \frac{1}{2E_\nu} U_L\hat M^2 U_L^\dagger & 
-B_+ U_L (\mu+id) U_L^T\\
-B_- U_L^* (\mu-id) U_L^\dagger & \frac{1}{2E_\nu} U_L^* \hat M^2 U_L^T
\ea \r) \,.
\ee
Note that $U_R$ in $H^D_\nu$ (\ref{HDnu}) is replaced by $U_L^*$ in the
Majorana case. Furthermore, we have to keep in mind that the matrices $\mu$
and $d$ are antisymmetric, i.e., 
\be
\mu^T = \mu^* = -\mu \,, \quad d^T = d^* = -d
\ee
in $H_M$. This equation formulates the well known properties of Majorana
neutrinos that only transition moments are non-zero and that $\mu$ and $d$ are
purely imaginary. The latter property allows to check immediately that
\be\label{JM}
J^\dagger H_M J = H_M^*
\ee
holds (compare with Eq.(\ref{J})).

Denoting the diagonalizing $2n \times 2n$ unitary matrix of $H_M$ by $U_M$,
the survival probabilities of Majorana neutrinos with negative and positive
helicities are given by
\be\label{PML}
P^M (\nu_{\alpha L}\to\nu_{\alpha L}) = 
\left| \, \sum_{j=1}^{2n} \left| U_{M\, -\alpha j} 
\right|^2 e^{-iE_j L} \, \right|^2
\ee
and
\be\label{PMR}
P^M (\nu_{\alpha R}\to\nu_{\alpha R}) = 
\left| \, \sum_{j=1}^{2n} \left| U_{M\, +\alpha j} 
\right|^2 e^{-iE_j L} \, \right|^2 \,,
\ee
respectively. The subscript $-$ of $U_M$ in Eq.(\ref{PML}) indicates
that the index $\alpha$ refers to the first $n$ lines of this matrix 
(negative helicity), whereas in Eq.(\ref{PMR}) the subscript $+$
indicates the use of the last $n$ lines (positive helicity).

The relation (\ref{JM}) allows to be more specific with respect to the
matrix $U_M$.\\[1mm]
\textbf{Lemma:} 
The matrix $U_M$ which diagonalizes $H_M$ can be chosen to be of the form
\[
U_M = \l(\ba{cr} A & B^*\\
B & -A^*\ea
 \r) \,.
\]
Furthermore, diagonalizing $H_M$ with this $U_M$ we obtain
\[ 
E_j = E_{n+j} \quad \mbox{for} \quad j = 1, \ldots, n. 
\]
\\[1mm]
\textbf{Proof of the lemma:}
Suppose we have an eigenvector $\psi$ of $H_M$, i.e.,
\be
H_M \psi  = E \psi \,.
\ee
Then with Eq.(\ref{JM}) we find that
\be
H_M J \psi^* = J \left( J^\dagger H_M J \right) \psi^* =
J H_M^* \psi^* = E J \psi^* 
\quad \mbox{and} \quad J \psi^* \perp \psi \,.
\ee
This observation allows to construct an orthonormal basis of
eigenvectors of $H_M$ in the following way. Starting with an
eigenvector $\psi_1$ normalized to unit length and with eigenvalue
$E_1$, then $J \psi^*_1$ is orthogonal to $\psi_1$ and has the same
eigenvalue. Next we can find a normalized eigenvector $\psi_2$ with
eigenvalue $E_2$ such that 
$\psi_2 \perp \{ \psi_1, J \psi^*_1 \}$. Then it is easy to show
that $J \psi^*_2$ is orthogonal to all three previously constructed
eigenvectors. We continue by finding $\psi_3$ with eigenvalue $E_3$
orthogonal to all four previously constructed eigenvectors and so on.
After $n$ steps, two orthonormal systems $\{ \psi_j \}_{j = 1, \ldots, n}$
and $\{ J \psi_j^* \}_{j = 1, \ldots, n}$ with the same eigenvalues
$E_1, \ldots, E_n$ are found which together form an orthonormal basis
of eigenvectors of $H_M$. Thus $U_M$ is given by
\be
U_M = 
\left( \psi_1 \cdots \psi_n \; J \psi^*_1 \cdots J \psi^*_n \right) \,,
\ee
which is of the form announced in the lemma. $\Box$

An immediate consequence of the lemma and of Eqs.(\ref{PML}) and
(\ref{PMR}) is the following theorem.\\[1mm]
\textbf{Theorem 1:} Without matter effects and in a constant magnetic
field the survival probabilities for left-handed and right-handed
Majorana neutrinos are equal, i.e., 
\[
P^M (\nu_{\alpha L}\to\nu_{\alpha L}) =
P^M (\nu_{\alpha R}\to\nu_{\alpha R}) =
\left| \sum _{j = 1}^n \left( |A_{\alpha j}|^2 + |B_{\alpha j}|^2
\right) e^{-i E_j L} \right|^2 \,.
\]

\section{ASYMMETRIES OF SURVIVAL PROBABILITIES}

Let us define an asymmetry 
\be\label{asym}
\Delta_\alpha^D = 
P^D (\nu_{\alpha L} \rightarrow \nu_{\alpha L}) -
P^D (\bar\nu_{\alpha R} \rightarrow \bar\nu_{\alpha R})
\ee
for every Dirac neutrino flavour or type $\alpha$. Note that we have
just discussed in Theorem 1 that the corresponding asymmetry for Majorana
neutrinos, 
\be\label{asymM}
\Delta_\alpha^M = 
P^M (\nu_{\alpha L} \rightarrow \nu_{\alpha L}) -
P^M (\nu_{\alpha R} \rightarrow \nu_{\alpha R}) \,,
\ee
vanishes with the assumptions in this theorem. 
If we set $\mu = d = 0$ and assume vanishing
matter effects, then we have vacuum oscillations and 
both asymmetries (\ref{asym}) and (\ref{asymM}) are zero.
Clearly, the asymmetry (\ref{asym}) is not a CP asymmetry 
due to a KM type of phase \cite{KM}. If it is non-zero it is 
because the presence of a magnetic field represents a CP-violating
situation. 

As a passing remark, matter effects would induce such an asymmetry for
the same reason: background matter is not CP-invariant or, in other words,
neutrinos and antineutrinos interact differently with matter. The matter
potentials enter in the diagonal of the Hamiltonian matrices as follows:
\be
\begin{array}{rl}
(V_L, 0) & \mbox{for Dirac neutrinos,}\\
(-V_L, 0) & \mbox{for Dirac antineutrinos,}\\
(V_L, -V_L) & \mbox{for Majorana neutrinos,}
\end{array}
\ee
with the diagonal matrix $V_L = \sqrt{2} G_F\, \mathrm{diag} ( N_\alpha )$, 
where, for ordinary matter,
$N_e = n_e - n_n/2$, $N_{\mu,\tau} =  - n_n/2$, $N_s = 0$ and $n_e$ and $n_n$
are the electron and neutron densities, respectively.

However, there
is a fundamental difference between matter effects and the effects of MMs,
EDMs and a magnetic field: For $B_\pm = 0$, but matter effects becoming
important, one has $\Delta_\alpha^D = \Delta_\alpha^M \neq 0$ in general,
whereas with $V_L = 0$ but $B_\pm \neq 0$ and constant magnetic field one has 
$\Delta_\alpha^D \neq 0$ in general, as we will see, 
but $\Delta_\alpha^M = 0$.
 
In view of Theorem 1 and in order to elaborate under which conditions
Majorana and Dirac neutrinos behave differently, it is necessary to
study the asymmetry (\ref{asym}) in more detail. The following theorem
describes under which sufficient conditions the asymmetry is still zero.
\\[1mm]
\textbf{Theorem 2:} For the case of Dirac neutrinos, without matter
effects and with a constant magnetic field, if the electric and
magnetic moment matrices obey the proportionality $\mu = c d$ (or 
$d = c \mu$), where $c$ is a real number, then the survival
probabilities of a neutrino with flavour (type) $\alpha$ is equal to
the survival probability of the corresponding antineutrino, i.e., the
asymmetry (\ref{asym}) is zero.
\\[1mm]
\textbf{Corollary:} If $d=0$ or $\mu=0$ or $n=1$ the asymmetry
(\ref{asym}) is zero.\\[1mm]
\textbf{Proof of the theorem:}
We define (see Eq.(\ref{B}))
\be
e^{i\beta} (1 + i c) \equiv z = |z|e^{i\zeta} \,.
\ee
We can write $\tilde H_\nu$ (\ref{Hnutilde}) as
\be
\tilde H_\nu = \l(\ba{cc} \frac{1}{2E_\nu}\hat M^2 & 
 -Bz\mu\\
-Bz^*\mu & \frac{1}{2E_\nu}\hat M^2 \ea
 \r) \,.
\ee
The phase $\zeta$ can be removed by the unitary transformation
\be 
H_\nu^\prime \equiv
\left( \begin{array}{cc} 
\mathbf{1} & 0 \\ 0 & {\mathbf{1}} e^{i\zeta} 
\end{array} \right)
\tilde H_\nu 
\left( \begin{array}{cc} 
\mathbf{1} & 0 \\ 0 & {\mathbf{1}} e^{-i\zeta} 
\end{array} \right) = 
\left( \begin{array}{cc} 
\frac{1}{2E_\nu}\hat M^2 & -B|z|\mu \\
-B|z|\mu & \frac{1}{2E_\nu}\hat M^2 
\end{array} \right) = 
\left( \begin{array}{rr} 
H_1 & -H_2 \\ -H_2 & H_1
\end{array} \right) \,.
\ee
The individual block matrices $H_1$ and $H_2$ are independent
Hermitian matrices. 

We consider the eigenvector equations
\bea
(H_1 - H_2) X_j &=& E_j X_j \,, \nonumber\\
(H_1 + H_2) Y_j &=& E_{n+j} Y_j \,,
\eea
where $j = 1,\ldots,n$. The sets $\{ X_j \}_{j=1,\ldots,n}$ and 
$\{ Y_j \}_{j=1,\ldots,n}$ form orthonormal bases of 
eigenvectors of the Hermitian matrices $H_1 - H_2$ and $H_1 + H_2$, 
respectively. Therefore, the structure of $W_\nu$ is given by
\be
W_\nu = \frac{1}{\sqrt{2}} \l(\ba{cccccc} 
X_1 & \cdots & X_n & Y_1 & \cdots & Y_n \\
e^{-i \zeta} X_1 & \cdots & e^{-i \zeta} X_n & 
-e^{-i \zeta} Y_1 & \cdots & -e^{-i \zeta} Y_n \ea
 \r) \equiv
\l(\ba{cc} C & D \\
e^{-i \zeta} C & -e^{-i \zeta} D \ea
 \r) \,.
\ee
Furthermore, using the relations (\ref{U}) and the expressions (\ref{PDnu}) and
(\ref{PDnubar}) for the Dirac neutrino survival probabilities, Theorem 2
follows. $\Box$
   
Now we want to show that Theorem 2 describes an exceptional situation and 
that indeed in general the asymmetry $\Delta^D_\alpha$ is different from
zero. It is sufficient to see this in the case of vanishing neutrino masses in
$H^D_\nu$ (\ref{HDnu}) and $H^D_{\bar\nu}$ (\ref{nubar}).
Defining a matrix \cite{grimus}
\be\label{lambda}
\lambda = \mu - i d \,,
\ee
we notice that this is a completely general matrix which can be bidiagonalized
with unitary matrices $R$ and $S$:
\be\label{RS}
\lambda = R \hat \lambda S^\dagger 
\quad \mbox{with} \quad 
S = (x_1, \ldots,x_n )\,, \quad R = ( y_1, \ldots, y_n ) \,,
\ee
where $\hat\lambda$ is diagonal and positive and $\{ x_j \}_{j=1,\ldots,n}$
and $\{ y_j \}_{j=1,\ldots,n}$ are orthonormal bases. Dropping the neutrino
masses we get the Hamiltonian matrix
\be\label{Hl}
H^D_\nu = -B \left( \begin{array}{cc}
0 & e^{i\beta} S \hat \lambda R^\dagger \\
e^{-i\beta} R \hat \lambda S^\dagger & 0
\end{array} \right) \,,
\ee
which has the following eigenvectors:
\begin{eqnarray}
\phi_j = \frac{1}{\sqrt{2}} \left( \begin{array}{c}
x_j \\[1mm] e^{-i\beta} y_j      
\end{array} \right)
& \quad \mbox{with eigenvalue} \quad & -B \hat \lambda_j \,, \nonumber\\
\psi_j = \frac{1}{\sqrt{2}} \left( \begin{array}{c}
x_j \\[1mm] -e^{-i\beta} y_j \end{array}\right)
& \quad \mbox{with eigenvalue} \quad & \hphantom{-} B \hat \lambda_j \,.
\end{eqnarray}
These eigenvectors of the Hamiltonian matrix (\ref{Hl}) form the matrix
$W_\nu$ (\ref{W}), and thus with Eqs.(\ref{PDnu}), (\ref{PDnubar}) 
and (\ref{U}) we obtain
\begin{eqnarray}
P^D (\nu_{\alpha L}\to\nu_{\alpha L}) & = &
\left| \, \sum_{j=1}^n | x_{\alpha j} |^2 \cos (B \hat \lambda_j L)
\, \right|^2 \,, \nonumber \\
P^D (\bar\nu_{\alpha R}\to\bar\nu_{\alpha R}) & = &
\left| \, \sum_{j=1}^n | y_{\alpha j} |^2 \cos (B \hat \lambda_j L)
\, \right|^2 \,. \label{PDlambda}
\end{eqnarray}
From these expressions it is obvious that in general the probabilities
$P^D (\nu_{\alpha L}\to\nu_{\alpha L})$ and
$P^D (\bar\nu_{\alpha R}\to\bar\nu_{\alpha R})$
are different,
because the orthonormal bases $\{ x_j \}_{j=1,\ldots,n}$
and $\{ y_j \}_{j=1,\ldots,n}$ are independent of each other. 

To elaborate
this in more detail we consider the case of two neutrino flavours
($n=2$). Since we neglect here the neutrino masses all phases in $\lambda$
(\ref{lambda}) can be removed and the matrices $S$ and $R$ (\ref{RS}) 
are characterized by the angles $\theta$ and $\theta'$, respectively:
\be
S = \left( \begin{array}{rr} \cos \theta & -\sin \theta \\
                             \sin \theta &  \cos \theta 
           \end{array}                              \right) 
= (x_1, x_2) \,,
\quad
R = \left( \begin{array}{rr} \cos \theta' & -\sin \theta' \\
                             \sin \theta' &  \cos \theta' 
           \end{array}                                \right)
= (y_1, y_2) \,.
\ee
Inserting $x_j$ and $y_j$ into the survival probabilities (\ref{PDlambda}), it
is evident that
$P^D (\nu_{\alpha L}\to\nu_{\alpha L}) \neq 
P^D (\bar\nu_{\alpha R}\to\bar\nu_{\alpha R})$ holds as long as 
$\hat\lambda_1 \neq \hat\lambda_2$ and 
$\cos^2 \theta \neq \cos^2 \theta'$.

\section{MAJORANA NEUTRINOS AND $\lowercase{z}$-DEPENDENT MAGNETIC FIELDS}

Even if we assume that $V_L = 0$ but $B_x$, $B_y$ depend on $z$,
the survival probabilities for left-handed and right-handed Majorana
neutrinos will in general be different, i.e., Theorem 1 will not hold
anymore. This provides some obstacle for an application of the results
of this paper with the aim to find a way to distinguish between the Dirac and
Majorana nature of neutrinos. We briefly discuss the general
formalism for the case of $z$-dependent $H_M$. One needs to find $2n$
linearly independent solutions of the differential equation
\be\label{diff}
i \frac{d \varphi(z)}{dz} = H_M(z) \varphi(z) \,.
\ee
The survival probabilities
depend now also on the locations $z_0$ of the neutrino source and $z_1$ of the
neutrino detection. Assuming to know a complete orthonormal set 
$\{ \varphi_j(z) \}_{j=1,\ldots,2n}$ of solutions of Eq.(\ref{diff}), 
we can formulate the transition and survival probabilities as  
\begin{eqnarray}
P^M (\nu_{\alpha L}(z_0)\to\nu_{\beta L}(z_1)) & = &
\left| \, \sum_{j=1}^{2n} \varphi_{-\alpha j}^*(z_0) 
\varphi_{-\beta j}(z_1) \, \right|^2
\,, \nonumber \\
P^M (\nu_{\alpha R}(z_0)\to\nu_{\beta R}(z_1)) & = &
\left| \, \sum_{j=1}^{2n} \varphi_{+\alpha j}^*(z_0) 
\varphi_{+\beta j}(z_1) \, \right|^2
\,, \label{PMz}
\end{eqnarray}
and similarly those with LR and RL transitions. 
We use the indices $-\alpha$ and $+\alpha$ to indicate the $n$ upper
and $n$ lower components of $\varphi$, respectively. 

If the matter potential can be neglected ($V_L = 0$), 
the relation (\ref{JM}) for the Majorana Hamiltonian matrix
holds, irrespective if $H_M$ depends on $z$ or not. It is then easy to check
that, given a complete orthonormal system of solutions 
$\{ \varphi_j(z) \}_{j=1,\ldots,2n}$, then
$\{ J\varphi_j^*(-z) \}_{j=1,\ldots,2n}$ is a complete orthonormal set of
solutions of the differential equation (\ref{diff}) with $H_M(z)$ replaced by
$H_M(-z)$. This suggests that with a Hamiltonian matrix $H_M$ symmetric in $z$
one can arrive at a situation where Theorem 1 is still valid. Indeed, 
using Eq.(\ref{PMz}) and replacing $\varphi_j(z)$ by $J \varphi_j^*(-z)$,
one arrives at the following relations for
the Majorana transition and survival probabilities.\\[1mm]
\textbf{Theorem 3:}
\be
V_L = 0, \; H_M(z) = H_M(-z) \quad \Rightarrow \quad
P^M (\nu_{\alpha L}(z_0)\to\nu_{\beta L}(z_1)) =
P^M (\nu_{\beta R}(-z_1)\to\nu_{\alpha R}(-z_0)) \,,
\ee
and, therefore, with $z_1 = -z_0$, we have
\be
P^M (\nu_{\alpha L}(-z_1)\to\nu_{\alpha L}(z_1)) =
P^M (\nu_{\alpha R}(-z_1)\to\nu_{\alpha R}(z_1)) \,.
\ee
Therefore, Theorem 1 holds also for a $z$-dependent magnetic field, provided
it is symmetric between the neutrino source and detection points.

Note that from Eq.(\ref{PMz}) by exchanging complex conjugation within the
absolute values, for any $V_L$  and $z$-dependent magnetic field, 
we obtain the relations
\begin{eqnarray}
P^M (\nu_{\alpha L}(z_0)\to\nu_{\beta L}(z_1)) & = &
P^M (\nu_{\beta L}(z_1)\to\nu_{\alpha L}(z_0)) \,, \nonumber \\
P^M (\nu_{\alpha R}(z_0)\to\nu_{\beta R}(z_1)) & = &
P^M (\nu_{\beta R}(z_1)\to\nu_{\alpha R}(z_0)) \,, 
\end{eqnarray}
which express CPT invariance. However, they do not allow to relate the
survival probabilities for different helicities and are, therefore, 
not useful in our context.

\section{SUMMARY}

In this paper we have studied the survival probabilities of neutrinos
and antineutrinos which possess magnetic moments and electric dipole
moments and propagate in magnetic fields. In particular, given a
neutrino flavour (type) $\alpha$, we have
studied the difference $\Delta^{D,M}_\alpha$ (\ref{asym},\ref{asymM})
between the survival probabilities for $\nu_{\alpha L}$ and 
$\stackrel{\scriptscriptstyle (-)}{\nu}_{\alpha R}$.
The bar on $\nu_{\alpha R}$ refers to Dirac neutrinos, without bar a
Majorana neutrino is understood. It is well known that 
matter effects lead to non-zero asymmetries $\Delta^{D,M}_\alpha$, but
without electromagnetic neutrino interactions one has
$\Delta^D_\alpha = \Delta^M_\alpha$ expressing the fact that with
neutrino oscillations one cannot distinguish between Dirac and
Majorana neutrinos.

Here we have considered the opposite situation: we have
assumed that matter effects are negligible but neutrinos have MMs and
EDMs and propagate in magnetic fields. Assuming a \emph{constant} magnetic
field, we have shown that in this case we have $\Delta^M_\alpha = 0$
(Theorem 1), but $\Delta^D_\alpha \neq 0$ in general. However, if the
MM and EDM matrices are proportional to each other in the case of Dirac
neutrinos, then the asymmetry of survival probabilities is zero as well,
i.e., $\Delta^D_\alpha = 0$ (Theorem 2). What happens if the magnetic
field is not constant along the neutrino path between source and
detector? In general, this will result in $\Delta^M_\alpha \neq 0$ for
Majorana neutrinos, but if the transverse magnetic field is symmetric
with respect to the center $z_c$ of the line connecting source and detector 
($B_{x,y}(z_c + z) = B_{x,y}(z_c - z)$) one still has 
$\Delta^M_\alpha = 0$ (Theorem 3).

The observations summarized here indicate a fundamental difference in the
behaviour of Dirac and Majorana neutrinos with respect to magnetic
fields, which is a consequence
of the fact that the Hermitian MM and EDM matrices $\mu$ and $d$ have
to be antisymmetric for Majorana neutrinos. 
Further research is necessary to see if
the results of this paper lead to realistic possibilities to
distinguish between the Dirac and Majorana nature of neutrinos.

\acknowledgements
One of us (Balaji) gratefully acknowledges the hospitality at the 
Institute for Theoretical Physics, University of Vienna, Austria, 
where this work was initiated.

\end{document}